\documentclass[twocolumn,aps,prb]{revtex4} 
\usepackage{graphics,amsmath,graphicx} 

\begin{document} 
\title{Polaronic features in the optical properties 
of the Holstein-$t$-$J$ model} 
\author{E. Cappelluti} 

\affiliation{SMC Research Center and ISC, INFM-CNR, v. dei Taurini 19, 
00185 Rome, Italy,\\ 
Dipartimento di Fisica, Universit\`a ``La Sapienza'', 
P.le A. Moro 2, 00185 Rome, Italy} 

\author{S. Ciuchi} 

\affiliation{Istituto Nazionale di Fisica della Materia and 
Dipartimento di Fisica\\ 
Universit\`a dell'Aquila, 
via Vetoio, I-67010 Coppito-L'Aquila, Italy} 

\author{S. Fratini} 

\affiliation{Institut N\'eel - CNRS \& Universit\'e Joseph Fourier \\ 
BP 166, F-38042 Grenoble Cedex 9, France} 

\begin{abstract} 
We derive the exact solution for the optical conductivity 
$\sigma(\omega)$
of one hole in the Holstein-$t$-$J$ model in the framework of dynamical 
mean-field theory (DMFT). 
We investigate the magnetic and phonon features 
associated with polaron formation as a function of the 
exchange coupling $J$, of the electron-phonon interaction $\lambda$ 
and of the temperature. Our  solution  
directly relates the features of the optical conductivity to the 
excitations in the single-particle spectral function, revealing 
two distinct mechanisms of closing and filling of the optical pseudogap 
that take place upon varying the microscopic parameters. 
We show that 
the optical absorption  at the polaron crossover is characterized by a 
coexistence of a magnon peak at low frequency and a broad polaronic 
band at higher frequency.
An analytical expression for $\sigma(\omega)$
valid in the polaronic regime is  presented.
\end{abstract} 
\date{\today} 
\maketitle

\section{Introduction}

The problem of a single hole in the $t$-$J$ model interacting also 
with the lattice degrees of freedom has recently attracted a notable 
interest in connection with the physical properties 
of the high-$T_c$ cuprates. \cite{mish1,rosch1,rosch3,gunn,prev} 
In particular, in parent   
and strongly underdoped compounds, angle resolved 
photoemission spectroscopy (ARPES) reveals a low energy peak 
whose dispersion is well described by the $t$-$J$ model, while its 
anomalously large broadness has been ascribed to incoherent multi-phonon 
shake-off processes.\cite{shen,shen2,mish1,rosch3} 
A similar interplay between electron-electron and electron-phonon 
interactions should in principle be reflected 
in the optical conductivity spectra. 
As a matter of fact, the most remarkable features observed in the underdoped 
region  are 
an ubiquitous mid-infrared (MIR) peak at $\approx 0.5$ eV, 
and a weaker peak around $\approx 0.1$ eV, the latter being more 
strongly doping and temperature dependent.\cite{MIR-review,MIR-review2} 
Several interpretations have been proposed for the origin of these features, 
including midgap or impurity states, 
\cite{impurityMIR1,impurityMIR2} 
charge/spin stripes,\cite{stripesMIR,stripesMIR2} 
polaronic 
excitations,\cite{polaronMIR,polaronMIRb,polaronMIR2,polaronMIR3,polaronMIR4} 
and the interaction with the 
antiferromagnetic background.\cite{impurityMIR2}

The first proposal of the $t$-$J$ model as a suitable basis 
to discuss the optical spectra of the cuprates 
was advanced by Zhang and Rice \cite{zr} 
who observed that the $1/\omega$ behavior\cite{timusk} of the optical 
absorption above  $\approx 0.5$ eV could be naturally 
associated to the incoherent motion in an antiferromagnetic background. 
Successively, the optical conductivity of the $t$-$J$ model 
has been investigated in detail using several 
techniques, such as exact-diagonalization,\cite{dagotto} 
analytical approximations\cite{tikofsky,bang,kyung_tJ,jackeli} 
and dynamical mean-field theory (DMFT).\cite{strack,logan,jarrell,haule} 
However, and in spite of the above discussed relevance of 
the electron-phonon coupling, the optical conductivity 
of the $t$-$J$ model {\em in the presence} of the lattice degrees of freedom 
has not been thoroughly investigated. 
Numerical calculations based on exact-diagonalization 
of finite clusters were employed for instance 
in Ref. \onlinecite{feshke} to evaluate $\sigma(\omega)$. 
Alternatively, the optical conductivity 
was calculated analytically  in Ref. \onlinecite{kyung} based 
on a non-crossing Born approximation, which is however unable to describe the 
polaron formation.

In this paper we present  results for the optical conductivity 
of a single hole in the Holstein-$t$-$J$ model obtained in the 
framework of dynamical mean-field theory. 
One-hole spectral properties at zero temperature were discussed in a 
previous publication
where antiferromagnetic correlations were shown to
enhance the effects of the electron-phonon coupling.\cite{cc}
A similar result was found also
in the antiferromagnetic phase of the Holstein-Hubbard model
using DMFT techniques.\cite{StJAF}
A serious drawback of DMFT, which is obtained as the exact solution of the 
lattice problem in the limit of infinite dimensions, is that the magnetic 
background is treated in a classical way. 
This, together with the fact that Trugman loops are negligible  in infinite 
dimensions, 
prevents the possibility 
to account for  coherent hole-propagation, which is related 
to the spin-flip fluctuations. As a consequence, no Drude peak 
can be observed in the optical conductivity. 
On the other hand, the incoherent contributions of $\sigma(\omega)$ 
are mainly dominated by {\em local} properties, 
such as the local electron-phonon scattering and 
spin-string excitations within the magnetic polaron, which 
are well captured by this approach.\cite{cc}

Bearing the above limitations in mind,  the aim of the present work is thus 
to focus on the incoherent part of the optical conductivity 
of a single-hole and to investigate in detail its features 
in the different physical regimes of the Holstein-$t$-$J$ model.
The dependence 
of the optical spectra on the microscopic parameters  is 
analyzed with special attention to the intermediate coupling region, 
where the interplay between magnetic and lattice degrees of freedom 
is strongest. 
We show that a crucial role is played 
by the formation of the lattice polaron, which drives a transfer 
of spectral weight towards higher frequencies,  opening a pseudogap 
at low frequencies. 
Conversely, starting from the polaronic phase,  two different 
mechanisms can be clearly identified 
as being responsible for the 
disappearance of the pseudogap: 
$i$) reducing the effective exchange energy scale 
suppresses the positive feedback of magnetism on polaron formation and 
can lead to a {\em closing} of the pseudogap as the system crosses back to 
the non-polaronic regime; 
$ii$) increasing the temperature, 
which does not alter the lattice/magnetic interplay, 
leads to a {\em filling} of the pseudogap more similar to what is 
expected in purely polaronic models. 
In the immediate vicinity of 
the polaron crossover, the spectra are characterized by a coexistence 
of a magnon peak at low frequency and a broad polaronic band at higher 
frequency, which closely resembles the experimental situation observed 
in the cuprates.

On theoretical grounds, the definition of the optical conductivity 
of a  single hole is a delicate matter which needs 
particular care. 
We provide an analytical derivation which generalizes 
the results of 
Refs. \onlinecite{logan,fratini} to the Holstein-$t$-$J$ model. 
This approach permits us to identify the role 
of the different one-particle properties on the 
optical conductivity. 
Comparison with numerical data 
is also discussed, showing a good agreement between our 
findings and exact diagonalization results.

The paper is organized as follows. In Sec. \ref{s-model} we discuss 
the exact solution in infinite dimensions 
for the one-hole Green's function of the Holstein-$t$-$J$ model 
at finite temperature. Results for the 
one-particle spectral features are discussed in Sec. \ref{s-spectral}. 
An analytical expression for the optical conductivity $\sigma(\omega)$
is derived 
in Sec. \ref{s-opt} where we also investigate
the different polaronic features and their dependence on 
the microscopic parameters. In Sec. \ref{s-sc} we present a further
simplified expression for $\sigma(\omega)$ valid in the lattice
polaron regime. The main results 
are briefly summarized in Sec. \ref{s-summary} where also
the consequences of including spin fluctuations (here neglected)
is also discussed.
Finally,
a detailed derivation of the analytical expression for the 
one-particle spectral function and the optical conductivity 
is reported in the Appendices.

\section{Holstein-$t$-$J$ model in infinite dimensions} 
\label{s-model} 

In the following we consider the case of a single hole in an 
antiferromagnetic (AF) background 
interacting with local Holstein phonons. 
Using the linear spin-wave approximation\cite{martinez,ramsak,marsiglio} 
and neglecting terms that vanish in the limit of 
large coordination number $z \gg 1$, we can write 
the Hamiltonian  as:\cite{cc} 
\begin{eqnarray} 
H&=&\frac{t}{2\sqrt{z}}\sum_{\langle ij \rangle} 
\left(h_j^\dagger h_i a_j + {\rm h.c.}\right) 
+\frac{J(1-2x)}{2}\sum_i a_i^\dagger a_i +\nonumber \\ 
&+&g\sum_i 
h_i^\dagger h_i 
\left(b_i+b_i^\dagger\right) 
+\omega_0 \sum_i b_i^\dagger b_i. 
\label{ham} 
\end{eqnarray} 
Here $a^\dagger$ is the creation operator for  boson {spin} defects, 
$h^\dagger$ is the single spinless hole operator and 
$x=\langle a^\dagger a \rangle$ 
represents the density of spin defects which is finite 
at nonzero temperature. 
Note that in the thermodynamical limit the presence of a single hole 
does not affect the magnetic state, 
which can be thus evaluated (in the $z \gg 1$ limit) 
in the absence of spin dynamics. 
The density of spin defects can 
be obtained from the magnetization $m=1-2x$ via the Curie-Weiss equation: 
\begin{eqnarray} 
m=\tanh \left(\frac{\beta Jm}{4}\right), 
\label{curie} 
\end{eqnarray} 
which defines a N\'eel temperature $T_{\rm N}=J/4$. 
Concerning the electron-lattice interaction, 
we shall mainly focus on the adiabatic  regime $\omega_0\ll t$ which is 
relevant to the experimental systems of interest. 
In this regime, a dimensionless electron-phonon coupling constant can 
be defined as $\lambda=g^2/\omega_0 t$, the polaron energy 
in units of the hopping integral.

An exact solution for the thermodynamical 
and the one-particle spectral properties of Eq. (\ref{ham}) 
at $T=0$ was obtained in Ref. \onlinecite{cc} 
in terms of 
a continued fraction. In order to derive the optical conductivity, 
the one-particle Green's function must be 
generalized to finite temperature, 
which involves the following steps: 
$i$) 
one has to allow for thermally  excited phonons; 
$ii$) 
the presence of thermally excited spin defects requires 
the introduction of a ``spin-resolved'' Green's function, to 
distinguish hole excitations created on sites with/without spin 
defects; 
$iii$) finally, 
the reduced magnetization 
introduces an effective exchange coupling $\tilde{J}=Jm < J$. 
The details of a formal derivation of the one-particle propagator 
at finite temperature are reported in Appendix \ref{app-green-T}; 
we summarize here the main results. 

Following Ref. \onlinecite{logan}, we define 
$\bar{G}_{i,0}(\omega)=\bar{G}_i(\omega)$ 
as the Green's function 
for one hole created on a site {\em in the absence} of spin defects. 
A careful analysis (see Appendix \ref{app-green-T}) shows that 
the Green's function $\bar{G}_{i,1}(\omega)$ 
for a hole created on a site with a  spin defect is simply 
$\bar{G}_{i,1}(\omega) =\bar{G}_i(\omega+J)$. 
In addition, at finite temperature the Green's function 
$\bar{G}_i(\omega)$ itself is defined as a thermal average 
over the phonons: 
\begin{eqnarray} 
\bar{G}(\omega) 
=\frac{1}{Z_{\rm ph}} 
\sum_n \mbox{e}^{-\beta n \omega_0} 
\bar{G}^{nn}(\omega+n\omega_0), 
\label{g-gnn} 
\end{eqnarray} 
where $\bar{G}^{nn}(\omega)$ represents the propagation of 
one hole created on a site with $n$ excited phonons and
$Z_{\rm ph}$ is the single-site phonon partition function
$Z_{\rm ph}=1/(1-\mbox{e}^{-\beta \omega_0})$.
Following Ref. \onlinecite{polarone}, we can derive a self-consistent 
expression for $\bar{G}^{nn}$ in terms of a continued fraction. 
We can write: 
\begin{eqnarray} 
\bar{G}^{nn}(\omega) 
= 
\frac{1}{ 
{\cal G}^{-1}(\omega-n\omega_0) 
-\Sigma_{\rm em}^n(\omega)-\Sigma_{\rm abs}^n(\omega)}, 
\label{fracGnn} 
\end{eqnarray} 
where 
\begin{widetext} 
\begin{eqnarray} 
\Sigma_{\rm em}^n(\omega) 
&=& 
\frac{(n+1)g^2}{\displaystyle 
{\cal G}^{-1}(\omega-(n+1)\omega_0)- 
\frac{\displaystyle   (n+2)g^2}{\displaystyle 
{\cal G}^{-1}(\omega-(n+2)\omega_0)- 
\frac{\displaystyle (n+3)g^2}{\displaystyle 
\ldots 
} 
} 
}, 
\label{Sigma_em} \\ 
\Sigma_{\rm abs}^n(\omega) 
&=& 
\frac{ng^2}{\displaystyle 
{\cal G}^{-1}(\omega-n\omega_0)- 
\frac{\displaystyle   (n-1)g^2}{\displaystyle 
{\cal G}^{-1}(\omega-(n-1)\omega_0)- 
\frac{\displaystyle (n-2)g^2}{\displaystyle 
\ldots 
} 
} 
}. 
\label{Sigma_abs} 
\end{eqnarray} 
\end{widetext} 
Carrier motion and exchange interactions are taken into account 
by the bath propagator 
\begin{equation} 
\label{bathG} 
{\cal G}^{-1}(\omega)=\omega-\Sigma_t(\omega), 
\end{equation} 
where the ``hopping'' self-energy is defined as 
\begin{equation} 
\Sigma_t(\omega)= 
\frac{t^2}{4} 
\left[ 
(1-x)\bar{G}_j(\omega-\tilde{J}/2) 
+x \bar{G}_j(\omega+\tilde{J}/2) 
\right]. 
\label{Sigma_t} 
\end{equation} 
Finally, the spin-defect averaged Green's function, which is 
the physically probed quantity in photoemission experiments, is obtained as: 
\begin{equation} 
G(\omega) 
= 
(1-x) \bar{G}(\omega) + 
x \bar{G}(\omega+\tilde{J}/2). 
\label{Gtot} 
\end{equation} 
Note that the factors $(1-x)$ and $x$ in front of 
$\bar{G}_j(\omega-\tilde{J}/2)$ and $\bar{G}_j(\omega+\tilde{J}/2)$ 
account for the probability 
of a site to be, respectively, free or populated by a spin defect. 
Note also that the Green's function $\bar{G}(\omega)$ 
appearing in Eqs. (\ref{bathG},\ref{Sigma_t}) is a phonon averaged quantity, 
so that the solution of Eqs. (\ref{g-gnn})-(\ref{Gtot}) 
involves the simultaneous self-consistency of all of the $\bar{G}_i^{nn}$, 
which is much more computationally expensive than 
the solution at zero temperature. 

It is easy to check that 
in the  absence of electron-phonon interaction 
Eq. (\ref{Gtot}) recovers the thermal Green's function 
for the pure $t$-$J$ model as defined 
in Eq. (2.12) by Stumpf and Logan.\cite{logan} 
On the other hand, it should be stressed that 
the present solution, 
although it formally recovers the results for the pure Holstein model 
in the limit $\tilde{J} \rightarrow 0$ (paramagnetic case) 
[Eqs. (40)-(42) of Ref. \onlinecite{polarone}], 
is still described by a  purely local 
Green's function $G_{i,j}(\omega)=\delta_{i,j}G(\omega)$ 
due to the assumption of a classical (although disordered) 
spin background. The main drawback of this assumption is thus 
that no coherent dispersive peak is obtained in this framework 
even in the $J \rightarrow 0$ limit of the $t$-$J$ Holstein model, 
and, consequently, no Drude peak appears in the 
optical conductivity. 
Notwithstanding this limitation 
the present approach, as we will show below, 
is still quite able to reproduce in more than qualitative agreement 
the incoherent features of both the one-particle Green's function 
and of the optical conductivity. 

\section{One-hole spectral properties} 
\label{s-spectral} 

Before discussing the optical conductivity, 
let us briefly present our results for the one-particle 
spectral function in the Holstein-$t$-$J$ model at finite temperature. 
Previous approaches have   
focused separately on thermal effects either in the pure Holstein or 
in the pure $t$-$J$ model. 
The temperature evolution of the hole 
spectral function $\rho(\omega)=-(1/\pi)\mbox{Im}G(\omega)$ 
for the $t$-$J$ model in the infinite 
dimensional limit has been analyzed 
in Ref. \onlinecite{logan}. At 
$T=0$, it consists of a series of $\delta$-function magnon peaks 
whose  distribution  reflects the 
strength of the magnetic polaron: for sufficiently 
large $J/t$ the profile is rapidly decaying with energy (reminiscent 
of the string picture of small magnetic polarons) whereas 
for small $J/t$ it acquires a more symmetric 
shape, reducing to a semi-circular function 
in the limit $J/t \rightarrow 0$. 
The effect of a non-zero temperature within this context 
is to broaden each $\delta$-function with 
a bandwidth $W_x$ which is ruled by the thermal spin defect 
probability $x$. We shall term this effect the {\em intrinsic magnetic 
broadening}. Such intrinsic magnetic broadening $W_x$ is however 
exponentially small for $T/T_{\rm N}\ll 1$, and it becomes 
significant only close to $T_{\rm N}$. A semi-circular shape is 
recovered also in the paramagnetic phase $T\geq  T_{\rm N}$, where 
$\tilde{J}/t=0$.

As the electron-lattice coupling is turned on, each magnon peak splits 
into several sub-peaks spaced by $\omega_0$, reflecting the dressing 
of the hole by phononic excitations. 
In this context the strength of the electron-phonon coupling 
rules not only the number of phonon satellites 
but also its spectral weight profile. 
Just as in the pure Holstein model, while the number of phonon peaks 
is quite small in the weak coupling regime, 
in the polaronic state a large number of phonon satellites appear 
with a characteristic Gaussian profile. 
The envelope of the phononic fine structure has  a spread which 
is governed by the energy associated with the lattice 
fluctuations: it is given by $\sqrt{\lambda \omega_0 t}=g$ in the 
quantum limit, and increases $\sqrt{2 \lambda T t}$ as $T\gtrsim 
\omega_0/2$. \cite{hohu}

\begin{figure}[t] 
\begin{center} 
\includegraphics[width=7.8cm,clip=]{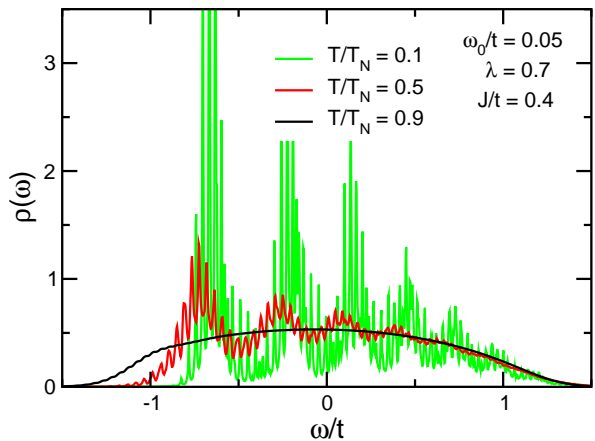} 
\end{center} 
\caption{(color online) 
Temperature evolution of the total spectral function 
$\rho(\omega)=-(1/\pi)\mbox{Im}G(\omega)$ 
in the polaronic regime. Microscopical parameters: 
$\lambda=0.7$, $\omega_0/t=0.05$, 
$J/t=0.4$ and $T/T_{\rm N}=0.1, 0.5, 0.9$ (corresponding to 
$T/\omega_0=0.2, 1, 1.8$). 
Numerical calculations 
have been done with a small imaginary frequency part $\eta=0.007$.} 
\label{f-sf} 
\end{figure}

In Fig. \ref{f-sf} we show the temperature evolution of a typical 
polaronic spectral function, 
at $\lambda=0.7$, $\omega_0/t=0.05$ 
and $J/t=0.4$. 
Throughout the paper, when not specified, we shall take the 
hopping matrix element $t$ as the energy unit. 
The temperatures  considered here are 
$T/T_{\rm N}=0.1, 0.5, 0.9$,  corresponding to 
$T/\omega_0=0.2, 1, 1.8$. 
At the lowest temperature ($T/T_{\rm N}=0.1$) 
a phononic fine structure can be clearly seen, 
superimposed on the magnon peaks. 
The width of each phononic peak is due to the intrinsic magnetic broadening 
$W_x$ described in Ref. \onlinecite{logan}. 
It is exponentially small at 
this temperature, so that a small Lorentzian broadening 
$\eta=0.007$ has been introduced for clarity. 
On the other hand, the  spread of the multiphonon 
structure gives rise to an overall width to the magnon peaks that 
is in good agreement with the expected value $g=0.19$.

Upon increasing the temperature to $T/T_{\rm N}=0.5$, 
two different effects are visible. 
First, the width $W_x$ of each of the fine peaks increases 
due to the intrinsic magnetic broadening, 
leading to a much smoother curve 
(this effect actually 
overcomes the small Lorentzian broadening $\eta$ introduced previously). 
Also, the overall spread of the multiphonon profiles 
increases due to the thermal phonon fluctuations, as expected for 
$T\gtrsim \omega_0/2$. 
Finally, at  $T/T_{\rm N}=0.9$,  the 
system is so close to the paramagnetic phase that 
neither the phonon peaks nor the 
magnon  structure can be resolved.

At this point, it is useful 
to comment about the effects of coherent hole propagation, that are implicitly 
neglected in our approach. 
These would induce a finite dispersion of 
order $\sim J$ to the lowest energy  magnon 
peak.\cite{mish1,feshke,kyung,martinez,kane,marsiglio} 
It is clear that such dispersion would be visible only 
at sufficiently low temperatures and at moderate electron-phonon 
coupling strengths, 
when the energy scale $J$ is smaller than both the intrinsic magnetic 
broadening and the Gaussian phonon spread, whereas in the opposite case 
it will presumably be hidden below a featureless background. 
These considerations  give further support to the present DMFT approach 
in the polaronic and/or high temperature  regime, where neglecting 
the coherent hole propagation would not affect significantly 
the spectral properties. 
As we shall see below, this is even more true 
for what regards the finite-frequency 
optical conductivity, where any dispersive  peak would 
be convolved in any case with high-energy featureless structures.

\section{Optical conductivity} 
\label{s-opt}

The evaluation of the optical conductivity 
for a single hole is a delicate matter, which is only partly 
simplified in the context of DMFT 
due to the absence of vertex corrections.\cite{khurana} 
A controlled procedure is derived in terms of an expansion of 
the inverse fugacity at finite temperature, performing 
the limit $\mu \rightarrow -\infty$ to enforce the thermodynamically 
vanishing particle density. 
We can thus define the optical conductivity per hole, 
$\sigma(\omega)=\lim_{n_h\rightarrow 0} \sigma(\omega;n_h)/n_h$, 
which is a finite quantity and which presents the same features 
as in the dilute (but finite) hole density limit. 
Applying this formalism one derives 
a similar expression as obtained in Refs. \onlinecite{zr,logan}, 
here adapted to take into account  the electron-phonon interaction. 
Leaving once more the technical details in Appendix \ref{app-sigma}, 
we report here the main results. The 
optical conductivity per hole is expressed as 
\begin{eqnarray} 
\sigma(\omega) 
&=& 
\frac{t^2\pi(1-\mbox{e}^{-\beta\omega})}{4\omega } 
\int d\Omega\, \bar{\rho}^{\rm w}(\Omega)\nonumber \\ 
&&\hspace{-13mm}\times 
\left[x \bar{\rho}(\omega+\Omega+\tilde{J}/2) 
+(1-x)\bar{\rho}(\omega+\Omega-\tilde{J}/2)\right], 
\label{sigma} 
\end{eqnarray} 
where 
\begin{eqnarray} 
\bar{\rho}(\omega) 
& = & 
-\frac{1}{\pi} \mbox{Im}\bar{G}(\omega), 
\label{rozz} 
\end{eqnarray} 
and 
\begin{eqnarray} 
\bar{\rho}^{\rm w}(\omega) 
& =& 
\frac{\mbox{e}^{-\beta \omega}\bar{\rho}(\omega)} 
{\int d\omega\, \mbox{e}^{-\beta \omega} \bar{\rho}(\omega)}. 
\label{wDOS} 
\end{eqnarray} 
The last line defines a ``weighted spectral function'', 
$\bar{\rho}^{\rm w}(\omega)$, 
which represents thermally excited states and 
plays an important role in the determination 
of the optical conductivity.

Let us  remark that, 
although Eqs. (\ref{sigma})-(\ref{wDOS}) are formally analogous 
to those for the 
pure $t$-$J$ model, the electron-phonon interaction appears implicitly 
in them 
in the evaluation of the local Green's function $\bar{G}(\omega)$. 
Note also that, 
in the paramagnetic limit $\tilde{J}\rightarrow 0$, 
Eqs. (\ref{sigma})-(\ref{wDOS}) do not recover the results of the 
Holstein model,\cite{fratini} because 
Eq. (\ref{sigma}) involves the convolution of two {\em local} rather than 
{\bf k}-{\em dependent} propagators. As discussed above, 
this is due to the classical treatment of the spin degrees of freedom 
which does not allow for coherent transport, so that 
no Drude peak is recovered in the present analysis. 
This, however, has only a minor influence on the finite-frequency optical 
conductivity, which is dominated by 
local incoherent excitations.

\begin{figure}[t] 
\begin{center} 
\includegraphics[width=6.5cm,clip=]{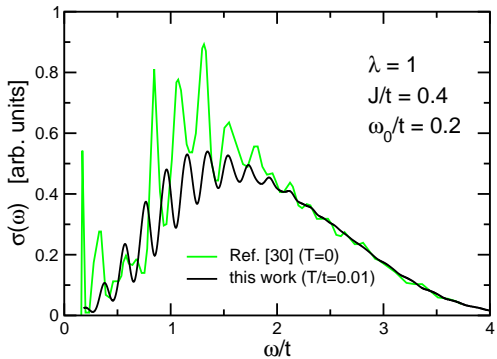} 
\end{center} 
\caption{(color online)
Comparison between the optical conductivity $\sigma(\omega)$ 
obtained by our DMFT 
solution and Lanczos diagonalization in two dimensions 
on a finite cluster (Ref. \onlinecite{feshke}, arbitrarily scaled). 
A Gaussian broadening $\Delta=3\omega_0/5$ 
for  $\sigma(\omega)$ has been employed in our DMFT analysis 
(see text for details).} 
\label{f-feshke} 
\end{figure} 

In order to assess the validity of the present treatment, 
we compare 
in Fig. \ref{f-feshke} the optical conductivity of the Holstein-$t$-$J$ 
model in infinite dimensions, 
as described by Eqs. (\ref{sigma})-(\ref{wDOS}), 
with numerical calculations using Lanczos diagonalization 
for a single hole in the 2D Holstein-$t$-$J$ model 
on a $\sqrt{10}\times\sqrt{10}$ 
cluster.\cite{feshke} 
For technical reasons (see discussion below), 
DMFT data are averaged with a Gaussian filter of amplitude
$\Delta=3\omega_0/5$ such that phonon resonances are  still well separated. 
Microscopic values are  $\lambda=1$, $J/t=0.4$, $\omega_0/t=0.2$, 
and $T=0$ (for Ref. \onlinecite{feshke}) 
and $T=0.01 t =0.1T_{\rm N}$ for the present results. 
These values correspond to a case where 
the lattice/magnetic polaron is formed 
and incoherent contributions to the optical conductivity 
are indeed dominant. 
The good agreement of the overall shape 
confirms the feasibility of our analysis 
to investigate the finite frequency 
optical conductivity.

\subsection{Technical details} 
\label{tech}

As discussed in the previous Section, 
the one-hole spectral function consists of a set of narrow 
bands whose width $W_x$ is controlled by the intrinsic magnetic 
broadening. This intrinsic width vanishes at low temperature due to 
the absence of spin-wave dispersion, 
making the direct evaluation of 
Eqs. (\ref{sigma})-(\ref{wDOS}) quite hard to perform. 
Special care is thus needed in order to calculate numerically 
the optical conductivity.

\begin{figure}[t] 
\begin{center} 
\includegraphics[width=8.0cm,height=9.2cm,clip=]{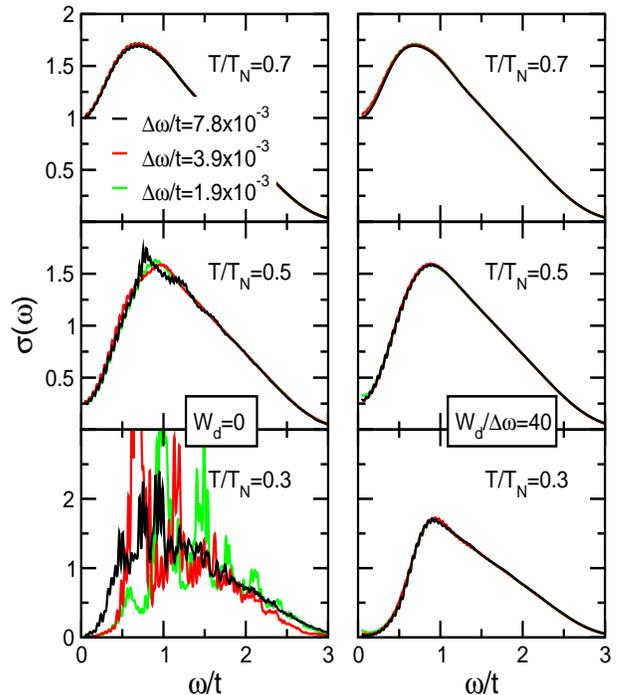} 
\end{center} 
\caption{(color online) 
Left panels: 
Gaussian averaged optical conductivity as a function 
of the sampling mesh $\Delta\omega$ for different temperatures. 
Right panel: the corresponding Gaussian averaged optical conductivity 
in the presence of a disorder-induced bandwidth 
$W_d/\Delta\omega=40$ as described by our scaling procedure. The 
microscopic parameters are $\lambda=0.7$, $\omega_0/t=0.05$ and 
$J/t=0.4$.  } 
\label{filtro} 
\end{figure}

We shall be mainly interested in the features of the optical 
conductivity that are related to the electron-lattice 
coupling. We shall therefore  retain the details of the spectra on 
the scale of the phonon frequency $\omega_0$. 
From the practical point of view, the primary object of our analysis 
will be a Gaussian average of the optical conductivity $\sigma(\omega)$ 
with a Gaussian filter of amplitude $\Delta=3\omega_0/5$.\cite{notagaux} 
Note that, even though 
the Gaussian average preserves the total spectral weight of 
the original set of data, the accuracy of the final result 
will be limited by the finite frequency sampling $\Delta\omega$. 
In explicit terms, if the spacing 
$\Delta\omega$ is larger than the intrinsic peak-width $W_x$, 
the data sampling will probe 
the spectral features in a random way, yielding a highly inaccurate 
result for both the  shape and spectral weight of the optical conductivity. 
This is shown in the left panels 
of Fig. \ref{filtro} where we plot the dependence of the 
(Gaussian averaged) optical conductivity on the sampling 
spacing $\Delta\omega$ for different temperatures. 
At high temperature $T/T_{\rm N}=0.7$ the intrinsic magnetic broadening 
$W_x$ is large enough so that both the shape and the total 
spectral weight of the optical conductivity are well captured even with 
a relatively large mesh ($\Delta\omega/t = 7.8\times 10^{-3}$). 
At lower temperature $T/T_{\rm N}=0.5$ 
however $W_x$ becomes so small that 
a much finer mesh ($\Delta\omega/t = 1.9\times 10^{-3}$) 
is needed in order to get accurate results. 
At $T/T_{\rm N}=0.3$, finally, 
no convergence  is achieved 
even for the finest 
sampling mesh considered in this paper, $\Delta\omega/t = 1.1\times 10^{-4}$ 
(in Fig. \ref{filtro}, for graphical reasons, we plot 
curves only up to $\Delta\omega/t = 1.9\times 10^{-3}$).
Clearly, since the 
intrinsic peak-width $W_x$ vanishes exponentially at low $T$, the 
problem evidenced here cannot be solved by merely reducing the sampling mesh.

To overcome this difficulty we add to the system 
a small uncorrelated disorder with semielliptic distribution 
of amplitude $W_d$,\cite{strack,logan} 
which is able to yield a finite peak-width even in the zero temperature limit. 
We choose $W_d=40\Delta\omega$ 
to assure a sufficiently dense mesh 
for an accurate sampling. 
We then scale $W_d \rightarrow 0$ keeping fixed the ratio 
$W_d/\Delta\omega=40$ in order to approach the correct physical limit 
in the absence of disorder.
Results obtained with a finite $W_d$
are shown in the right panels of Fig. \ref{filtro}. 
No appreciable difference is visible  at high temperature 
$T/T_{\rm N}=0.7$ where the thermally driven magnetic broadening $W_x$ 
is large enough to guarantee the convergence even in the absence 
of disorder. 
On the other hand our procedure provides a clear improvement 
already at $T/T_{\rm N}=0.5$ where the convergence 
as function of the sampling spacing is more easily achieved 
in the presence of disorder (note that the converged 
results  coincide with the results for the most dense mesh 
in the {\em absence} of disorder, showing 
that no spurious structures appear due to our scaling procedure). 
Finally, for $T/T_{\rm N}=0.3$ the disorder scaling procedure 
is the only way to guarantee convergence of both the shape 
and the total spectral weight of the optical conductivity.

\subsection{Results} 

Using the above described procedure, 
we shall now  concentrate on the evolution of the optical conductivity in 
the adiabatic regime, fixing the ratio $\omega_0/t=0.05$. 
This value is qualitatively representative of the cuprates, 
where the half-bandwidth $t \approx 1.2$ eV and typical optical phonon 
frequencies $\omega_{\rm ph} \approx 60$ meV. 
This regime is also the most interesting one from the theoretical 
point of view, since in this case 
the interplay between lattice and spin degrees of freedom has 
its most dramatic effects. 

\begin{figure}[t] 
\centering   
\includegraphics[width=7.5cm,clip=]{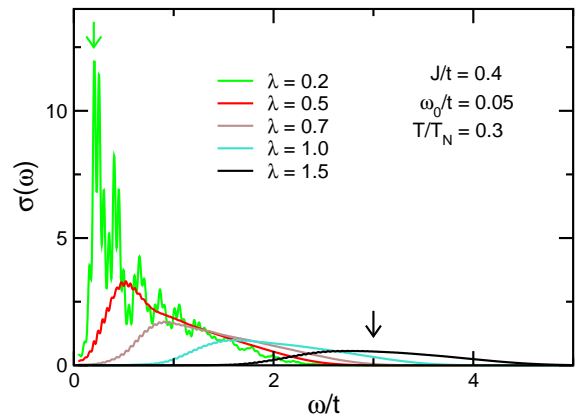} 
\caption{(color online) Optical conductivity versus $\lambda$ 
across the polaron 
crossover, at fixed $J=0.4$, $\omega_0/t=0.05$ and 
$T/\omega_0=0.6$ ($T/T_N=0.3$). 
The two arrows at low and high energy mark respectively the 
first magnon peak at $\omega \approx J/2$ in the weak electron-phonon 
coupling limit $\lambda \ll 1$, and the position $\omega \approx 
2\lambda t$ of the broad polaronic absorption expected 
at $\lambda \gg 1$ (shown here for $\lambda=1.5$).} 
\label{fig:scanL} 
\end{figure} 

Fig. \ref{fig:scanL} shows the evolution of the optical absorption at 
fixed $J/t=0.4$ and low temperature $T/T_N=0.3$, upon varying the 
electron-lattice coupling.  At $\lambda=0.2$, the result is very 
reminiscent of the spectra calculated for the pure $t$-$J$ model in 
Ref. \onlinecite{logan}. It consists of a 
series of magnetic peaks, dominated by the sharp single-magnon peak located at 
$\omega\simeq J/2$ and rapidly decaying at higher frequency. 
In this weak-coupling regime $\lambda=0.2$, 
the electron-phonon interaction 
simply gives rise to a 
multi-phonon fine-structure with a Gaussian profile. 
Each magnon peak acquires thus a phonon-driven width 
without modifying however   
the overall distribution of spectral weight.

The main effect of increasing the electron-lattice coupling is a 
progressive shift of the spectral weight towards higher frequencies. 
This is an evidence of the formation of the lattice polaron, 
which occurs through a 
gradual crossover in the presence of a finite $\omega_0$. 
Increasing $\lambda$ also modifies the shape of 
the low-energy absorption edge, converting the sharp magnon-peak 
at $\lambda\to 0$ into a smoother Gaussian lineshape, typical of 
polaronic absorption. 
In the strong coupling regime ($\lambda \gg 1$), 
characteristic of a  small lattice polaron, 
the position  of the  maximum in the optical conductivity is expected 
to scale linearly as $\omega=2 \lambda t$. 
This is in good qualitative agreement with our data 
reported in Fig. \ref{fig:scanL} where however 
the effects of a finite hopping integral $t$ (Ref. \onlinecite{fratini}) 
and of the $1/\omega$ behavior at high frequency (see Ref. \onlinecite{zr}) 
result in a slight redshift of the maximum of the 
polaronic structure. 

\begin{figure}[t] 
 \centering   
\includegraphics[width=7.5cm,clip=]{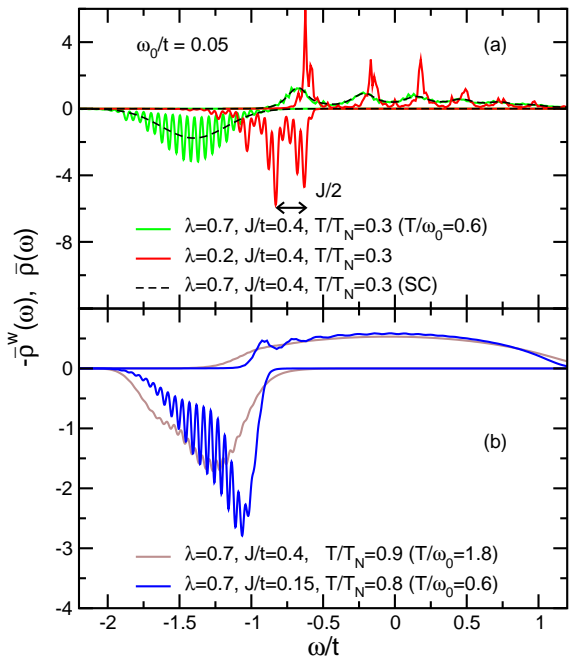} 
\caption{(color online) Spectral function $\bar{\rho}(\omega)$ 
and weighted spectral function $\bar{\rho}^{\rm w}(\omega)$ 
for the different cases reported in the legend. For better readability, 
the weighted spectral function $\bar{\rho}^{\rm w}(\omega)$ 
is reported on the negative axis. The dashed black line in panel (a) refers
to the strong coupling (SC) approximate formula as discussed
in Sec. \ref{s-sc}.} 
\label{fig:dostot} 
\end{figure}

The evolution of the optical conductivity with $\lambda$ 
can be understood by analyzing the building 
blocks of Eqs. (\ref{rozz},\ref{wDOS}).
In Fig. \ref{fig:dostot}(a), 
we report both the spectral function $\bar{\rho}$ and the weighted 
spectral function $\bar{\rho}^{\rm w}$ 
for two typical values $\lambda=0.2$  and $\lambda=0.7$. In the first 
case (dark red curve), there is no energy separation between the single-hole 
excitations in  $\bar{\rho}$ and the thermally excited states in 
$\bar{\rho}^{\rm w}$. The low-frequency gap in the optical absorption 
seen in Fig. \ref{fig:scanL} arises due to the explicit  shift of 
the spectral function by the quantity $\tilde{J}/2$ in 
Eq. (\ref{sigma}), representing the energy cost to create one 
spin-defect as the hole hops in 
the AF background. 
It is now interesting to compare these features with the results for 
$\lambda=0.7$ [light green curve in Fig. \ref{fig:dostot}(a)]. 
As we can see, the spectral functions $\bar{\rho}(\omega)$ 
for $\lambda=0.7$ and  $\lambda=0.2$ are qualitatively similar, 
the only major difference being 
the increased number of phonon satellites involved 
in each magnetic peak, reflecting the increased 
number of phonons in the polaron cloud. 
In order to understand the modification 
of the optical conductivity we  thus focus 
on the weighted spectral function 
$\bar{\rho}^{\rm w}$. The latter undergoes a much more drastic change 
reflecting in an explicit way the signature of polaron formation. 
Of particular relevance is the shift of $\bar{\rho}^{\rm w}$ to much 
more negative energies which characterizes the lattice trapping. 
Note also that 
the magnetic peaks 
that are clearly visible at $\lambda=0.2$ are completely  washed out at 
$\lambda=0.7$, merging into a broad polaronic-like spectrum 
centered at higher ``binding'' energies. 
This change in the nature of the thermally excited states 
is at the origin of both the opening of a polaronic pseudogap 
and of the smoothening of the features observed in  $\sigma(\omega)$.

\begin{figure}[t] 
 \centering   
\includegraphics[width=7.5cm,clip=]{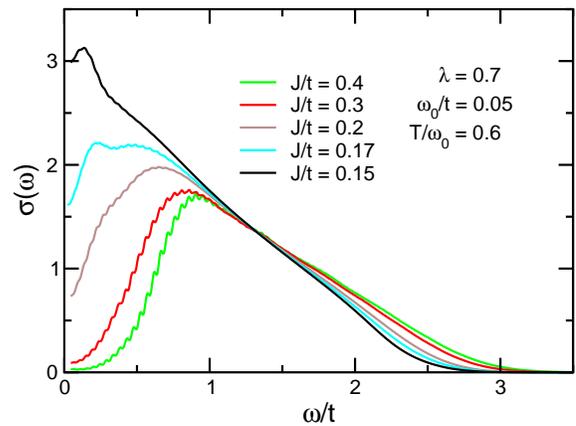} 
 \caption{(color online) Optical conductivity across the polaron 
crossover, as varying $J$ for fixed $\lambda=0.7$, $\omega_0/t=0.05$ and 
fixed temperature $T/\omega_0=0.6$. Note that this fixed temperature 
corresponds to different $T/T_{\rm N}$ for $J/t=0.4, 0.3, 0.2, 0.17, 0.15$, 
respectively $T/T_{\rm N}=0.3, 0.4, 0.6, 0.7, 0.8$.} 
\label{fig:scanJ} 
\end{figure} 

A qualitatively similar evolution is observed upon varying $J$, which 
gives a clear illustration of the 
positive interplay between magnetic and lattice polaron effects. This 
is shown in Fig. \ref{fig:scanJ} where we report the 
optical conductivity for different values of $J$ at constant $\lambda=0.7$ 
and at the same temperature $T/\omega_0=0.6$ as in Fig. \ref{fig:scanL} 
(note that this corresponds 
to different $T/T_{\rm N}$ as $T_{\rm N}$ scales with $J$). 
Remarkably, even though $\lambda$ is kept constant, 
reducing the magnetic exchange from the initial value $J=0.4$ 
leads to a gradual loss of the lattice polaronic features 
and to the undressing of the hole from multi-phonon excitations. 
This results in a shift 
of spectral weight from the polaronic peak to a magnon peak 
at lower frequency. 
The magnon peak is already visible as a shoulder  in the 
data of Fig. \ref{fig:scanJ}  at $J/t=0.3$ and $J/t=0.2$, and it 
emerges more clearly at lower values of $J$. 
At $J=0.17$, in particular, both the magnetic peak and the broad 
polaronic band are visible in  the absorption spectrum. 
The featureless nature of the absorption curves at low $J/t$
can be ascribed to the increasing disorder of the AF 
environment as $J$ diminishes and $T/T_N$ increases.\cite{logan}

Once again, the comparison of 
$\bar{\rho}^{\rm w}$ and $\bar{\rho}$ for $J/t=0.4$ [light green curve 
in \ref{fig:dostot}(a)] and for $J/t=0.15$ 
[dark blue curve in \ref{fig:dostot}(b)] 
allows to visualize in a simple way the loss of the 
polaronic features. 
Even in this case the most relevant quantity is the weighted spectral function 
$\bar{\rho}^{\rm w}$ whose main  excitations, for $J=0.15$, are shifted 
to much less negative energies than for $J/t=0.4$, closing the 
gap between $\bar{\rho}^{\rm w}$ and $\bar{\rho}$. Note in addition that, 
although less evident than at $J/t=0.4$, a magnetic peak in 
the spectral function $\bar{\rho}$ is still visible even for 
$J/t=0.15$. The convolution of $\bar{\rho}$ with $\bar{\rho}^{\rm w}$ 
gives rise to the small magnetic peak at 
$\omega \simeq \tilde{J}/2$ observed in the optical conductivity.

\begin{figure}[t] 
\centering   
\includegraphics[width=7.5cm,clip=]{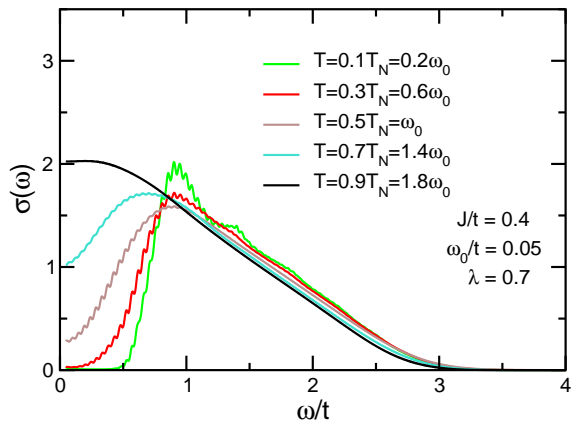} 
\caption{(color online) Temperature evolution of the 
optical conductivity for $\lambda=0.7$, $J/t=0.4$ and 
$\omega_0/t=0.05$.} 
\label{fig:scanT} 
\end{figure}

In Fig. \ref{fig:scanT} we report  the evolution of the optical spectra 
versus 
temperature in the polaronic regime at $\lambda=0.7$, $\omega_0/t=0.05$ and 
$J/4=0.4$. 
Although Fig. \ref{fig:scanT} can look at a first glance 
quite similar to Fig. \ref{fig:scanL}, 
no shift of the peak of $\sigma(\omega)$ 
is observed here. Instead,  
there is a progressive {\em filling} of the low frequency gap with
temperature that is 
quite similar to what is observed in the pure Holstein 
model.\cite{fratini} This 
can be pointed out again by the comparative study 
of $\bar{\rho}$ with $\bar{\rho}^{\rm w}$. 
For $T/T_{\rm N}=0.9$ [light brown curve in \ref{fig:dostot}(b)] 
the weighted spectral function $\bar{\rho}^{\rm w}(\omega)$ 
still presents  an extended peak at negative energy 
$\omega/t \approx -1.3$, signalizing that the lattice polaron is not 
completely destroyed. However, the broadening 
of $\bar{\rho}^{\rm w}(\omega)$ is now significantly enhanced as 
$T \gtrsim \omega_0$. This feature, along with a similar broadening of 
the spectral function $\rho(\omega)$ due to the approaching of 
the paramagnetic limit $T/T_{\rm N} \rightarrow 1$
(see Sec. \ref{s-spectral}), 
leads to 
a significant overlap of the two spectral functions 
and to a continuous filling of the gap in the optical conductivity 
as the temperature is increased. 

\section{Strong coupling formula} 
\label{s-sc}

In the previous section  we have discussed the  technical difficulties
involved in the calculation of the optical conductivity.
Even adopting the proposed scaling procedure (cf. Sec. \ref{tech}), 
the computational cost can still  be  quite demanding, 
especially in the strong coupling lattice polaron regime where 
the number of phonons to be taken into account scales as 
$\alpha^2=\lambda t/\omega_0$. 
In this section we present a simple analytical formulation 
of the optical conductivity which 
is valid precisely in the adiabatic ($\omega_0\rightarrow0$) 
small polaron regime and which 
involves almost no numerical effort. 

The basis of this simplified formulation is the 
polaronic nature of the weighted spectral function $\bar{\rho}^{\rm w}$ 
pointed out in  Fig. \ref{fig:dostot}. 
We have seen indeed that 
at moderate values 
of the electron-lattice coupling, just above the polaron 
crossover, the function $\bar{\rho}^{\rm w}$ becomes essentially independent 
of the exchange term $J$.
A closer look shows that in the lattice polaron regime
the spectral function $\bar{\rho}^{\rm w}$ can be
well approximated with the result
of the atomic limit $t\to 0$,\cite{Mahan}
\begin{equation} 
\label{rhoWSC} 
\bar{\rho}^{\rm w}(\omega) 
= 
\frac{1}{\sqrt{ 2 \pi s^2}}
\exp\left[-\frac{(\omega+2\lambda t)^2}{2s^2}\right],
\end{equation} 
where $s=g/\sqrt{\tanh(\omega_0/2T)}$.
This function for the case $\lambda=0.7$,
$J/t=0.4$, $\omega_0/t=0.05$, $T/T_{\rm N}=0.3$
is shown in Fig. \ref{fig:dostot}a compared with the full
numerical solution for the same parameters.
The good agreement can be ascribed to the atomic nature 
of the thermally induced excitations, which do not hybridize with the 
hopping continuum described by the self-energy Eq. (\ref{Sigma_t}) 
(an equivalent result was already pointed out in Ref. \onlinecite{fratini}).
Eq. (\ref{rhoWSC}) provides thus a simple analytical expression
for $\bar{\rho}^{\rm w}(\omega)$ which can be employed in
Eq. (\ref{sigma}) for the evaluation of
the optical conductivity.

A simple analytical approximation can
also be derived for  the one-hole spectral function $\bar{\rho}(\omega)$
involved in Eq. (\ref{sigma}).
At this level, we are mainly interested in the overall shape
of the optical conductivity $\sigma(\omega)$, disregarding
the multi-phonon fine structure that will be anyway washed out at 
large $\lambda$ as shown in Fig. \ref{fig:scanL}.
In this perspective
we can consider formally the limit $\omega_0 \rightarrow 0$
and use the approach devised in 
Refs. \onlinecite{polarone,hohu} as well as in Ref. \onlinecite{rosch1} 
to treat the adiabatic limit. 
In particular,
the sum over the discrete energy levels in Eq. (\ref{g-gnn})
can be replaced by an integration over
Gaussianly distributed local random energies
$\nu$ which account for 
the thermal/quantum fluctuations of the phonon field. 
The local propagator thus becomes\cite{polarone,hohu,rosch1} 
\begin{equation} 
\label{eq:avprop} 
\bar{G}(\omega)=\int d\nu \frac{P(\nu)}{{\mathcal G}^{-1}(\omega)-\nu}, 
\end{equation} 
where $P(\nu)=(1/\sqrt{ 2 \pi s^2})\mbox{e}^{-\nu^2/2s^2}$
is a Gaussian function with the same
variance as in Eq. (\ref{rhoWSC} ). 
Self-consistency is achieved by employing
Eqs. (\ref{bathG})-(\ref{Sigma_t}).
The comparison between the approximate formula
for $\bar{\rho}(\omega)$ valid in the polaronic regime and the
full numerical solution is also  shown in
Fig. \ref{fig:dostot}a for $\lambda=0.7$,
$J/t=0.4$, $\omega_0/t=0.05$, $T/T_{\rm N}=0.3$.
Note that, although
the bath propagator ${\mathcal G}(\omega)$ does not depend
explicitly on the local random energies $\nu$, it still depends
parametrically on $\lambda$ and $J$ through the self-consistency
conditions Eqs. (\ref{bathG})-(\ref{Sigma_t}), so that it still retains
the relevant lattice and magnetic spectral structures.

\begin{figure}[t] 
 \centering 
\includegraphics[width=7.5cm,clip=]{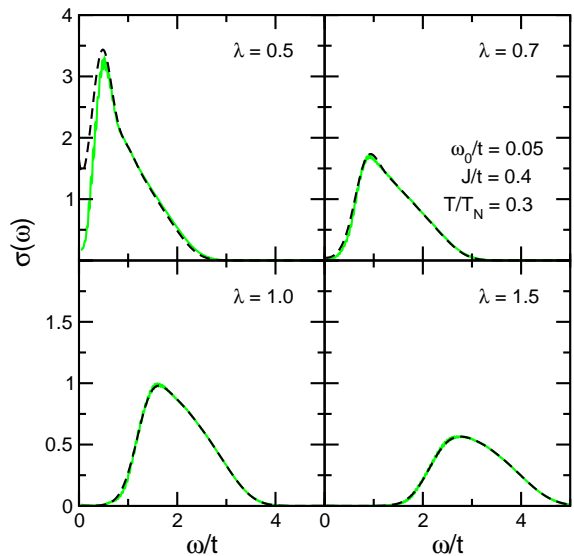} 
\caption{(color online) Comparison of the optical conductivity
evaluated from the full numerical solution (light green lines)
and by means of the strong coupling formula
(dashed black lines).
The agrement
is accurate already at moderate values of $\lambda$.} 
\label{fig:SC} 
\end{figure}

Once the approximate analytical expressions for
the spectral functions 
$\bar{\rho}$ and $\bar{\rho}^{\rm w}$ are provided,
we can now simply evaluate the
optical conductivity by using 
Eq. (\ref{sigma}). Let us stress that the numerical solution
of the approximate expressions for $\bar{\rho}$ and $\bar{\rho}^{\rm w}$,
defined in Eqs. (\ref{rhoWSC})-(\ref{eq:avprop}) requires
a remarkably smaller computational cost than the full numerical solution,
especially in the lattice polaronic regime
where Eqs. (\ref{rhoWSC})-(\ref{eq:avprop}) are valid and where
the computational cost of the full numerical solution is highest.
The comparison of the approximate formula with
the full numerical solution as function
of the electron-phonon coupling constant $\lambda$ is
illustrated in Fig. \ref{fig:SC}.
The agreement is excellent in the small polaron regime $\lambda \gtrsim 0.7$
and it is still quite satisfactory even for moderate
electron-phonon coupling $\lambda = 0.5$.

\section{Summary and conclusions} 
\label{s-summary}

In this work, we have provided an analytical treatment 
for the optical conductivity $\sigma(\omega)$ 
of one hole in the Holstein-$t$-$J$ model at finite temperature, 
in the limit of infinite dimensions. 
In this context a dynamical mean-field solution can be derived 
where the local self-consistent problem is solved exactly. 
Due to intrinsic limitations 
enforced in  infinite dimensions, we are not able 
to account for the coherent propagation of a single hole 
in the antiferromagnetic background. Nevertheless, we have 
shown how the incoherent optical processes, related to local excitations 
(Einstein phonons and local spin-flips) are well described in 
our approach, as confirmed by the comparison 
with numerical results obtained by Lanczos 
diagonalization. 

Our main aim has been to investigate the 
incoherent features of $\sigma(\omega)$ 
in an intermediate coupling region where the positive 
interplay between the magnetic 
and lattice degrees of freedom is more relevant, sustaining the 
formation of a spin-lattice polaron. 
In this context we have studied the evolution of polaronic features 
in the optical conductivity as function of the different microscopic 
parameters, as the electron-phonon coupling $\lambda$, 
the temperature $T$ and the effective exchange energy $\tilde{J}$. 
We remind that $\tilde{J}$ does not represent the bare exchange energy 
but rather a mean-field-like Weiss exchange coupling 
which depends on the local magnetization $m$. 
In the cuprates, for example, this parameter can be tuned 
by varying the hole doping starting from the parent AF phase. 
We have shown that the role of the electron-phonon coupling 
is twofold. 
On one hand it rules the formation of the lattice polaron, 
changing the incoherent part of $\sigma(\omega)$ 
from a typical spin-polaron spectrum, characterized by magnetic 
peaks and by an optical gap at $J/2$, 
to a broad lattice-polaron-like shape located at higher frequency, 
from which the magnetic peaks are essentially washed out. 
On the other hand it also tunes the amount of quantum lattice fluctuations, 
reflected in the emergence of 
multi-phonon satellite peaks. This fine structure 
survives also when the lattice polaron is destroyed 
at small $\lambda$ and it gives rise to an effective broadening 
of the magnetic peaks which can be much larger than the intrinsic broadening 
driven by the thermal magnetic fluctuations. 

The present approach allows us to distinguish 
between two different mechanisms leading to a 
suppression of the polaronic pseudogap: 
$i$) at intermediate values of the electron-phonon coupling, 
the reduction of the effective 
exchange energy $\tilde{J}$ 
leads to a shift of spectral weight 
from high energy lattice polaronic features 
to low energy magnetic excitations. This  results in a 
{\em closing} of the pseudogap 
as the hole undresses from its lattice polaron cloud; 
$ii$) 
Conversely, increasing the temperature within the polaronic regime, 
gives rise to a {\em filling} of the pseudogap. 
Both mechanisms are actually observed in the optical spectra of the 
underdoped cuprates.\cite{onose}

As a final remark, we briefly comment on  the robustness of our results 
in physical systems where quantum spin fluctuations are present, 
allowing for coherent motion of the holes. 
As discussed above, one of the main effects 
is the emergence of a dispersive pole 
with small spectral weight in the one-particle spectral function. 
In the optical spectra, this 
gives rise to a Drude-like low-frequency response,   but 
it does not affect the high-frequency incoherent part 
(moreover the coherent spectral weight is strongly 
reduced when the electron-lattice coupling is turned on). 
Regarding the high-frequency part, 
the intrinsic dispersion  of the spin fluctuations is 
itself expected 
to smear the magnetic peaks. This would affect our results only in the weak 
electron-phonon coupling regime, where no lattice polaron is formed. 
In this case an intrinsic broadening of the magnetic peaks 
in the optical spectra due to the spin-fluctuation dispersion 
should be considered for a quantitative analysis. 
On the other hand, the effects of the dispersion of 
spin-fluctuations are expected to be barely visible in the 
lattice polaron regime, where the smearing due to the multi-phonon satellite 
structure around each magnetic peak dominates. 

\acknowledgments

We acknowledge C. Bernhard and
G. Sangiovanni for stimulating discussions.
E.C. and S.C. acknowledge also financial support from the
Research Program MIUR-PRIN 2005.

\appendix

\section{Finite temperature Green's function of the Holstein-$t$-$J$ model 
in infinite dimensions} 
\label{app-green-T} 

In this Appendix we provide a detailed derivation 
of the Green's function of one hole in the Holstein-$t$-$J$ model 
at finite temperature in the infinite dimensional limit. 
A formal derivation for the pure $t$-$J$ model 
was discussed in Ref. \onlinecite{logan}, while the derivation 
for the full Holstein-$t$-$J$ model at $T=0$ was provided in 
Ref. \onlinecite{cc}. 
On this ground, here we limit ourselves to the 
derivation of the  finite temperature  self-consistent equations 
in terms of continued fractions.

Let us start by writing the Hamiltonian in Eq. (\ref{ham}) 
as $H=H_t+H_{\rm L}$, where $H_t$ represents the non-local hopping 
term while $H_{\rm L}$ contains all the other, purely local, contributions. 
We also define the Green's function as: 
\begin{widetext} 
\begin{eqnarray} 
G_i(t) 
&=& 
-i \theta (t) \frac{1}{Z^{\rm T}(N)} 
\sum_{\{n\}_i,\{s\}_i} 
\left\langle \{s\}_{j \neq i},\{n\}_{j \neq i};s_i,n_i \right| 
\mbox{e}^{-\beta H} 
\mbox{e}^{-iHt} 
c_i^\dagger 
\mbox{e}^{iHt} 
c_i 
\left|n_i,s_i=0;\{n\}_{j \neq i},\{s\}_{j \neq i} \right\rangle, 
\label{gdef} 
\end{eqnarray} 
\end{widetext} 
where $\left|n_i,s_i;\{n\}_{j \neq i},\{s\}_{j \neq i} \right\rangle$ 
denotes the state with $n_i$ phonons and $s_i$ spin defects 
($s_i=0$: no spin defect, $s_i=1$: spin defect) on the site $i$ 
and with a generic set $\{n\}_{j \neq i}$ of phonons 
and $\{s\}_{j \neq i}$ of spin defects 
on all the other sites. With these notations 
$c_i=h_i^\dagger$ when $s_i=0$ and 
$c_i=h_i^\dagger a_i$ when $s_i=1$. 
In addition, $Z^{\rm T}(N)$ is the total 
partition function of the system in the absence of holes, $N$ being 
the number of sites; in this case $H_t$ does not contribute and 
$Z^{\rm T}(N)$ factorizes into a phonon and a spin part as 
$Z^{\rm T}(N)=Z^{\rm ph}(N)Z^{\rm spin}(N)$. 
Each of them can be in addition factorized with respect to the 
site index, e.g. 
$Z^{\rm spin}(N)= Z^{\rm spin}_i\prod_{j\neq i} Z^{\rm spin}_j$. 
Similar considerations hold true for the exponential terms 
$\mbox{e}^{-\beta H}\mbox{e}^{-iHt}$ which apply 
on the states with no holes on the right side of Eq. (\ref{gdef}). Reminding 
$H_{\rm L}|n_i,s_i\rangle = n_i\omega_0+s_i \tilde{J}/2$, 
we can write, after few straightforward steps, 
in the Fourier space: 
\begin{widetext} 
\begin{eqnarray} 
G_i(\omega) 
&=& 
\sum_{n_i,s_i} 
\frac{\mbox{e}^{-\beta [n_i \omega_0 +s_i \tilde{J}/2]}} 
{Z^{\rm ph}_i Z^{\rm spin}_i} 
\sum_{\{n\}_{j \neq i}, \{s\}_{j \neq i}} 
\frac{\mbox{e}^{-\beta \sum_{j \neq i}[n_j \omega_0 +s_j \tilde{J}/2]}} 
{Z^{\rm ph}_{j \neq i} Z^{\rm spin}_{j \neq i}} 
\nonumber\\ 
&& 
\times 
\left\langle \{s\}_{j \neq i},\{n\}_{j \neq i};s_i,n_i \right| 
c_i^\dagger 
\frac{1}{\omega+n_i\omega_0+s_i\tilde{J}/2-H} 
c_i 
\left|n_i,s_i;\{n\}_{j \neq i},\{s\}_{j \neq i} \right\rangle. 
\label{gdefw} 
\end{eqnarray} 

This can be rewritten as 
\begin{eqnarray} 
G_i(\omega) 
&=& 
\frac{1}{Z^{\rm ph}} 
\sum_n \mbox{e}^{-\beta n \omega_0} 
G_i^{nn}(\omega+n\omega_0), 
\label{g-gnn-app} 
\end{eqnarray} 
where 
\begin{eqnarray} 
G_i^{nn}(\omega) 
&=& 
\sum_{s_i} 
\frac{\mbox{e}^{-\beta s_i \tilde{J}/2}} 
{Z^{\rm spin}_i} 
\sum_{\{n\}_{j \neq i}, \{s\}_{j \neq i}} 
\frac{\mbox{e}^{-\beta \sum_{j \neq i}[n_j \omega_0 +s_j \tilde{J}/2]}} 
{Z^{\rm ph}_{j \neq i} Z^{\rm spin}_{j \neq i}} 
\nonumber\\ 
&& 
\times 
\left\langle \{s\}_{j \neq i},\{n\}_{j \neq i};s_i,n_i \right| 
c_i^\dagger 
\frac{1}{\omega+s_i\tilde{J}/2-H} 
c_i 
\left|n_i,s_i;\{n\}_{j \neq i},\{s\}_{j \neq i} \right\rangle. 
\label{gnn-def} 
\end{eqnarray} 
Similarly, the factor 
$p_s=\mbox{e}^{-\beta s_i \tilde{J}/2}/Z^{\rm spin}_i$ 
defines the local population of spin defects which can be evaluated 
within the mean-field theory enforced by the infinite dimensional limit, 
namely $p_1=x$, $p_0=1-x$. 
We have thus: 
\begin{equation} 
G_i^{nn}(\omega) 
= 
(1-x) \bar{G}_{i,0}^{nn}(\omega) + 
x \bar{G}_{i,1}^{nn}(\omega), 
\end{equation} 
with 
\begin{eqnarray} 
\bar{G}_{i,s_i}^{nn}(\omega) 
&=& 
\sum_{\{n\}_{j \neq i}, \{s\}_{j \neq i}} 
\frac{\mbox{e}^{-\beta \sum_{j \neq i}[n_j \omega_0 +s_j \tilde{J}/2]}} 
{Z^{\rm ph}_{j \neq i} Z^{\rm spin}_{j \neq i}} 
\left\langle \{s\}_{j \neq i},\{n\}_{j \neq i};s_i,n_i \right| 
c_i^\dagger 
\frac{1}{\omega+s_i\tilde{J}/2-H} 
c_i 
\left|n_i,s_i;\{n\}_{j \neq i},\{s\}_{j \neq i} \right\rangle. 
\end{eqnarray} 

Finally, using the definition of $c_i$, one can see that 
$\bar{G}_{i,1}^{nn}(\omega)=\bar{G}_{i,0}^{nn}(\omega+\tilde{J}/2)$, 
and we can write 
\begin{equation} 
G_i^{nn}(\omega) 
= 
(1-x) \bar{G}_i^{nn}(\omega) + 
x \bar{G}_i^{nn}(\omega+\tilde{J}/2), 
\end{equation} 
where 
\begin{eqnarray} 
\bar{G}_i^{nn}(\omega) 
&=& 
\sum_{\{n\}_{j \neq i}, \{s\}_{j \neq i}} 
\frac{\mbox{e}^{-\beta \sum_{j \neq i}[n_j \omega_0 +s_j \tilde{J}/2]}} 
{Z^{\rm ph}_{j \neq i} Z^{\rm spin}_{j \neq i}} 
\left\langle \{s\}_{j \neq i},\{n\}_{j \neq i};n_i \right| 
\frac{1}{\omega-H} 
\left|n_i;\{n\}_{j \neq i},\{s\}_{j \neq i} \right\rangle. 
\label{bargnn-def} 
\end{eqnarray} 
\end{widetext} 
Here $\left|n_i;\{n\}_{j \neq i},\{s\}_{j \neq i}  \right\rangle$ 
denotes the state with $n_i$ phonons on  site $i$, 
$\{n\}_{j\neq i}$, $\{s\}_{j \neq i}$ being the 
phonon/spin configurations on the 
sites $j \neq i$ with {\em one hole} on the site $i$. 

We can write Eq. (\ref{bargnn-def}) using the short-hand notation: 
\begin{eqnarray} 
\bar{G}_i^{nn}(\omega) 
&=& 
\hat{P}_i 
\left\langle n_i \right| 
\frac{1}{\omega-H} 
\left|n_i\right\rangle, 
\label{bargnn} 
\end{eqnarray} 
where 
$|n_i\rangle\equiv |n_i;\{n\}_{j \neq i},\{s\}_{j \neq i}\rangle$, 
and where the operator 
\begin{eqnarray} 
\hat{P}_i 
\left\langle n_i \right| 
\ldots 
\left|n_i\right\rangle 
&=& 
\sum_{\{n\}_{j \neq i}, \{s\}_{j \neq i}} 
\frac{\mbox{e}^{-\beta \sum_{j \neq i}[n_j \omega_0 +s_j \tilde{J}/2]}} 
{Z^{\rm ph}_{j \neq i} Z^{\rm spin}_{j \neq i}} 
\nonumber\\ 
&& 
\hspace{-2cm}\times 
\left\langle \{s\}_{j \neq i},\{n\}_{j \neq i}\right| 
\ldots 
\left|\{n\}_{j \neq i},\{s\}_{j \neq i} \right\rangle, 
\end{eqnarray} 
denotes the average over the spin and phonon configurations 
on all the sites but $i$. 
The operator $\hat{P}_i$ represents a direct generalization 
of the quantity $P(s)$ introduced by Stumpf and Logan 
in Ref. \onlinecite{logan} to include the phonon degrees 
of freedom. 
It is also convenient to note that (for $j\neq i$): 
\begin{eqnarray} 
\hat{P}_i 
\left\langle n_i \right| 
\ldots 
\left|n_i\right\rangle 
&=& 
\hat{P}_i 
\sum_{n_j,s_j} 
\hat{P}_{ij} 
\frac{\mbox{e}^{-\beta [n_j \omega_0 +s_j \tilde{J}/2]}} 
{Z^{\rm ph}_j Z^{\rm spin}_j} 
\nonumber\\ 
&& 
\hspace{-2cm}\times 
\left\langle s_j,n_j,n_i\right| 
\ldots 
\left|n_i,n_j,s_j\right\rangle, 
\label{Pop} 
\end{eqnarray} 
where $\hat{P}_{ij}$ is defined in similar way as 
$\hat{P}_i$ as the average over the spin and phonon configurations 
on all the sites except $i$ and $j$.

Having introduced the necessary definitions, from now on we can 
follow the derivation in Ref. \onlinecite{cc} properly 
adapted to the finite temperature case. 
In particular we can introduce 
the local Green's function $\bar{g}_i^{nn}(\omega)$ defined 
as the atomic $t=0$ limit of Eq. (\ref{bargnn}). 
Note that in the atomic limit $\hat{P}_i=1$ so that 
\begin{eqnarray} 
\bar{g}_i^{nn}(\omega) 
&=& 
\frac{1}{\omega-n\omega_0}. 
\label{glocnn} 
\end{eqnarray} 
We can also generalize Eq. (\ref{bargnn}) 
for off-diagonal local phonon matrix elements: 
\begin{eqnarray} 
\bar{g}_i^{np}(\omega) 
&=& 
\hat{P}_i 
\left\langle n_i \right| 
\frac{1}{\omega-H_{\rm L}} 
\left|p_i\right\rangle, 
\end{eqnarray} 
whose analytical expression will be provided 
In Eq. (\ref{glocnm}). 

We can now employ the standard relation 
\begin{eqnarray} 
\frac{1}{\omega-H} &=& 
\frac{1}{\omega-H_{\rm L}}+ 
\frac{1}{\omega-H_{\rm L}}H_t 
\frac{1}{\omega-H_{\rm L}}\nonumber\\ 
&+& 
\frac{1}{\omega-H_{\rm L}}H_t 
\frac{1}{\omega-H_{\rm L}}H_t 
\frac{1}{\omega-H_{\rm L}}+\ldots, 
\end{eqnarray} 
which, on a classical spin background, gives rise to 
the retraceable path approximation. We have thus 
\begin{eqnarray} 
\bar{G}_i^{nm}(\omega) 
&=& 
\bar{g}_i^{nm}(\omega) 
-\sum_p 
\bar{g}_i^{np}(\omega) 
\bar{\Sigma}_j^{(p)_i}(\omega) 
\bar{G}_i^{pm}(\omega), 
\label{recursion} 
\end{eqnarray} 
where $\bar{\Sigma}_j^{(p)_i}(\omega)$ represents the dynamics 
of the hole after hopping on the neighboring site $j$. 
Since, in the leading term of a $1/d$ expansion, 
the hopping process and the further dynamics of the hole 
do not involve 
the phonon degrees of freedom on site $i$, such evolution 
occurs {\em in the presence} of $p$ phonons 
on the site $i$, which are reflected in an shift 
of the frequency argument, 
$\bar{\Sigma}_j^{(p)_i}(\omega)=\bar{\Sigma}_j(\omega-p\omega_0)$. 
Note that $\bar{\Sigma}_j(\omega)$ still contains the full thermal 
average on the phonons at site $j$, so that it is related 
to the thermally averaged Green's function defined in Eq. (\ref{g-gnn-app}). 
In addition, $\bar{\Sigma}_j(\omega)$ will depend also on the 
initial spin configuration $s_j$ at site $j$: 
hopping to a site $j$ free of spin defects 
will {\em create} a spin defect on the site $i$, while hopping to 
a site $i$ with a spin defect will restore the initial magnetic 
background at site $i$ {\em destroying} a spin defect at $i$. 
Using these considerations we have thus: 
\begin{eqnarray} 
\Sigma_j(\omega) 
&=& 
\frac{t^2}{4} 
\left[\sum_{s_j} p_{s_j} 
\bar{G}_j(\omega+(2s_j-1)\tilde{J}/2) \right] 
\nonumber\\ 
&&\hspace{-14mm} 
= 
\frac{t^2}{4} 
\left[ 
(1-x)\bar{G}_j(\omega-\tilde{J}/2) 
+x \bar{G}_j(\omega+\tilde{J}/2) 
\right], 
\label{Sigma_t-app} 
\end{eqnarray} 
and we can write Eq. (\ref{recursion}) in the compact form: 
\begin{eqnarray} 
\bar{G}^{nm}(\omega) 
= 
\bar{g}^{nm}(\omega) 
- 
\sum_p 
\bar{g}_i^{np}(\omega) 
\Sigma_t(\omega-p\omega_0) 
\bar{G}^{pm}(\omega), 
\end{eqnarray} 
where we have dropped the unnecessary site indices and 
we have added the index $\Sigma_t$ to denote a self-energy term 
arising from the hopping processes.

We can indeed employ now the usual procedure to obtain the electron-phonon 
self-energy in terms of a continued fraction. 
In particular, in addition to the diagonal elements 
of Eq. (\ref{glocnn}), we specify also the non-diagonal ones 
which read: 
\begin{eqnarray} 
\left[\bar{g}(\omega)\right]^{nm}= 
\left[\omega-n\omega_0\right]\delta_{n,m}+g X^{n,m}. 
\label{glocnm} 
\end{eqnarray} 

Using the standard derivation, we can thus write 
the Green's function $\bar{G}_i^{nn}(\omega)$ as 
\begin{eqnarray} 
\bar{G}^{nn}(\omega) 
= 
\frac{1}{ 
{\cal G}^{-1}(\omega-n\omega_0) 
-\Sigma_{\rm em}^n(\omega)-\Sigma_{\rm abs}^n(\omega)}, 
\end{eqnarray} 
where ${\cal G}^{-1}(\omega)=\omega-\Sigma_t(\omega)$, 
and 
\begin{widetext} 
\begin{eqnarray} 
\Sigma_{\rm em}^n(\omega) 
= 
\frac{(n+1)g^2}{\displaystyle 
{\cal G}^{-1}(\omega-(n+1)\omega_0)- 
\frac{\displaystyle   (n+2)g^2}{\displaystyle 
{\cal G}^{-1}(\omega-(n+2)\omega_0)- 
\frac{\displaystyle (n+3)g^2}{\displaystyle 
\ldots 
} 
} 
}, 
\end{eqnarray} 
and 
\begin{eqnarray} 
\Sigma_{\rm abs}^n(\omega) 
= 
\frac{ng^2}{\displaystyle 
{\cal G}^{-1}(\omega-n\omega_0)- 
\frac{\displaystyle   (n-1)g^2}{\displaystyle 
{\cal G}^{-1}(\omega-(n-1)\omega_0)- 
\frac{\displaystyle (n-2)g^2}{\displaystyle 
\ldots 
} 
} 
}. 
\label{Sabs} 
\end{eqnarray} 
\end{widetext}

\section{Optical conductivity} 
\label{app-sigma} 

In the previous Appendix we have derived an exact expression for 
the Green's function of a single hole in the Holstein-$t$-$J$ model 
in infinite dimensions. 
Here we investigate within the same framework the optical conductivity 
per hole, $\sigma(\omega)$, in the zero density limit. 
To this aim we provide an alternative derivation 
with respect to Ref. \onlinecite{logan}. 
The main advantage of the present approach is to deal at the same level 
with both the spin and phonon degrees of freedom, allowing thus 
for an immediate generalization of the $t$-$J$ model 
to the Holstein-$t$-$J$ model. 
As a result we obtain a final expression 
of the optical conductivity as a functional of 
the one-hole Green's function which is formally similar 
to the one of  Ref. \onlinecite{logan} but where 
the local Green's function takes into account 
the electron-phonon interaction.

The formal way to derive the optical conductivity $\sigma(\omega)$ 
for a single charge, 
is to consider the limit $\sigma(\omega)= 
\lim_{n_h\rightarrow 0} \sigma(\omega;n_h)/n_h$, 
where $\sigma(\omega;n_h)$, $n_h$ are quantities defined 
in the grand-canonical ensemble $\langle O \rangle 
= \sum_{N_h} \mbox{e}^{\beta \mu N_h} 
\mbox{\rm Tr}\left\{ O \right\}_{N_h}/Z_{\rm G.C.}$, 
and where $N_h$ it the total 
number of charges. The limit $n_h\rightarrow 0$ is enforced by 
expanding to lowest order in terms of the fugacity 
$z=\mbox{e}^{\beta \mu}$ ($\mu \rightarrow -\infty$). 
In this limit only the subsector $N_h=1$ survives both in 
$\sigma(\omega;n_h)$ and in $n_h$. 

Let us first consider the hole density, which we can write as: 
\begin{eqnarray} 
n_h=\frac{\mbox{e}^{\beta \mu}}{ Z^{\rm T}} 
\sum_{i,\alpha} 
\mbox{e}^{-\beta E_\alpha} 
\left\langle \alpha \left| 
h_i^\dagger h_i 
\right| \alpha \right\rangle, 
\label{ndef} 
\end{eqnarray} 
where 
$| \alpha \rangle$ is a complete set of eigenstates 
with eigenvalues $E_\alpha$ of one single hole (subspace $N_h=1$). 
Here, since the state $| \alpha \rangle$ must contain one hole 
at site $i$, the hole-number operator is simply defined 
as $\hat{N}_{h,i}=h_i^\dagger h_i$ without any spin defect. 
Inserting now into Eq. (\ref{ndef}) a complete set of eigenstates 
$|\gamma\rangle$ 
in the subspace $N_h=0$ (no hole) and a $\delta$-function, we obtain 
\begin{eqnarray} 
n_h 
&=& 
\mbox{e}^{\beta \mu} 
\int d\omega \mbox{e}^{-\beta \omega} 
\nonumber\\ 
&\times& 
\left[ 
\sum_{i,\alpha,\gamma} \frac{\mbox{e}^{-\beta E_\gamma}}{Z^{\rm T}} 
\left| \left\langle \alpha  \left| 
h_i^\dagger \right| \gamma \right\rangle \right|^2 
\delta(\omega-E_\alpha+E_\gamma) 
\right]. 
\label{n1} 
\end{eqnarray} 
Note that only the $|\gamma\rangle$ which do not contain any spin 
defect at site $i$ would contribute to Eq. (\ref{n1}), otherwise, 
after the hole $h^\dagger$ creation, we would end up with 
a state with one hole and one spin defect present, which is 
forbidden in the Hilbert space. 
We have thus: 
\begin{eqnarray} 
n_h 
&=& 
N \mbox{e}^{\beta \mu} p_0 
\int d\omega \mbox{e}^{-\beta \omega} 
\nonumber\\ 
&\times& 
\left[ 
\sum_{\alpha,\gamma} \frac{\mbox{e}^{-\beta E_\gamma}}{Z^{\rm T}} 
\left| \left\langle \alpha  \left| 
h_i^\dagger  \right| \gamma \right\rangle \right|^2 
\delta(\omega-E_\alpha+E_\gamma) 
\right], 
\label{n2} 
\end{eqnarray} 
where $p_0$ is the statistical probability to have 
a site with no spin defect and where 
now only the $|\gamma\rangle$ with no spin defect 
at the site $i$ are selected. 
The square bracket in Eq. (\ref{n2}) is just the local 
spectral function 
$\bar{\rho}(\omega)=-(1/\pi)\mbox{Im}\bar{G}_0(\omega)$, 
namely 
\begin{eqnarray} 
\bar{\rho}(\omega)= 
\sum_{\alpha,\gamma} \frac{\mbox{e}^{-\beta E_\gamma}}{Z^{\rm T}} 
\left| \left\langle \alpha  \left| 
h_i^\dagger  \right| \gamma \right\rangle \right|^2 
\delta(\omega-E_\alpha+E_\gamma), 
\label{rho} 
\end{eqnarray} 
[remind that $\bar{G}_0(\omega)=\bar{G}(\omega)$], 
so that we have 
\begin{eqnarray} 
n_h 
&=& N 
\mbox{e}^{\beta \mu}p_0 
\int d\omega \mbox{e}^{-\beta \omega} 
\bar{\rho}(\omega). 
\label{nfin} 
\end{eqnarray}

Let us turn now to the optical conductivity or, more precisely, 
to the current-current response function $\Pi(\omega)$ which 
is related to $\sigma(\omega;n_h)$ through the relation 
$\sigma(\omega;n_h)=-\mbox{Im}\Pi(\omega+i\delta)/\omega$. 
According to the previous argumentation, in the limit 
$n_h \rightarrow 0$ we can limit our analysis to the 
$N_h=1$ subspace and write: 
\begin{equation} 
\Pi(\tau)=-\frac{\mbox{e}^{\beta \mu}}{Z^{\rm T}} 
\mbox{Tr}\left\{ 
T_\tau \mathcal{J}(\tau)\mathcal{J} 
\right\}_{N_h=1}, 
\label{jj} 
\end{equation} 
where $\tau$ is the imaginary time in the Matsubara space and 
$\mathcal{J}$ is the current operator to be defined below. 

After usual manipulations we can write 
in the Fourier space: 
\begin{equation} 
\Pi(i\omega_m)= \mbox{e}^{\beta \mu} 
\sum_\alpha \frac{\mbox{e}^{-\beta E_\alpha}}{Z^{\rm T}} 
\left\langle \alpha \left| 
\mathcal{J} 
\frac{1-\mbox{e}^{-\beta(H-E_\alpha)}}{i\omega_m-H+E_\alpha} 
\mathcal{J} 
\right| \alpha \right\rangle, 
\label{jj3} 
\end{equation} 
where $\omega_m=2\pi m T$ are bosonic frequencies and where 
we remind that $|\alpha\rangle$ are eigenstates in the $N_h=1$ subspace.

The current density operator  can be written as 
$\mathcal{J}=(it/2\sqrt{z})\sum_{\langle i,j\rangle} 
\left( c_i^\dagger c_j-c_j^\dagger c_i\right)$, where we 
are summing explicitly on all possible directions (this prescription 
compensates for the 
well-known vanishing of the current-current response in infinite 
dimensions\cite{strack}).   
As discussed in Refs. \onlinecite{cc,martinez} and as appearing 
in Eq. (\ref{ham}), 
$c_i^\dagger c_j=h_i h_j^\dagger a_j$ if there is a spin defect 
on the site $j$ ($s_j=1$), while 
$c_i^\dagger c_j=a_i^\dagger h_i  h_j^\dagger$ if there is {\em no} 
spin defect on the site $j$ ($s_j=0$). 
Due to the classical nature of the magnetic background, 
it is easy to realize that a 
retraceable path approximation is enforced also 
in the current-current response function just as in the one-particle 
Green's function. We obtain thus: 
\begin{eqnarray} 
\Pi(i\omega_m) 
&=& \frac{t^2 \mbox{e}^{\beta \mu}}{4z} 
\sum_\alpha \frac{\mbox{e}^{-\beta E_\alpha}}{Z^{\rm T}} 
\nonumber\\ 
&&\times 
\sum_{\langle i,j \rangle} 
\left\langle \alpha \left| 
c_i^\dagger c_j 
\frac{1-\mbox{e}^{-\beta(H-E_\alpha)}}{i\omega_m-H+E_\alpha} 
c_j^\dagger c_i 
\right| \alpha \right\rangle. 
\label{jj4} 
\end{eqnarray}

As usual, 
we can now insert twice in Eq. (\ref{jj4}) the identity operator 
$\sum_\gamma |\gamma\rangle \langle \gamma |$, where $|\gamma\rangle$ 
are eigenstates in the subspace {\em without} any hole. 
We have thus: 
\begin{widetext} 
\begin{eqnarray} 
\Pi(i\omega_m) 
&=& 
\frac{t^2 \mbox{e}^{\beta \mu}}{4z} 
\sum_{\alpha,\gamma,\gamma'} 
\frac{\mbox{e}^{-\beta E_\alpha}}{Z^{\rm T}} 
\sum_{\langle i,j \rangle} 
\left\langle \alpha \left| c_j \right| \gamma \right\rangle 
\left\langle \gamma \left| 
c_i^\dagger 
\frac{1-\mbox{e}^{-\beta(H-E_\alpha)}}{i\omega_m-H+E_\alpha} 
c_i 
\right| \gamma' \right\rangle 
\left\langle \gamma' \left| c_j^\dagger \right| \alpha \right\rangle . 
\label{jj5} 
\end{eqnarray} 

Let us summarize the physical meaning of Eq. (\ref{jj5}). 
The eigenstate $|\alpha \rangle$ contains one hole 
at the site $j$, and Eq. (\ref{jj5}) describes the hopping 
of the hole to site $i$. 
If no spin defect is initially present at $i$ ($s_i=0$), this process 
involves the creation of a spin defect at $j$, and hence 
$c_i=h_i^\dagger$, $c_j^\dagger=a_j^\dagger h_j$. 
On the other hand, in the alternative case where 
a spin defect is initially present at $i$ ($s_i=1$), 
the hopping process destroys the spin defect at the 
site $i$, so that $c_i=a_i h_i^\dagger$ and $c_j^\dagger= h_j$. 

Let us consider for the moment the first case. 
We have thus: 
\begin{eqnarray} 
\Pi_0(i\omega_m) 
&=& 
\frac{t^2 \mbox{e}^{\beta \mu}}{4z} 
\sum_{\alpha,\gamma,\gamma'} 
\frac{\mbox{e}^{-\beta E_\alpha}}{Z^{\rm T}} 
\sum_{\langle i,j \rangle} 
\left\langle \alpha \left| h_j^\dagger a_j \right| \gamma \right\rangle 
\left\langle \gamma \left| 
h_i 
\frac{1-\mbox{e}^{-\beta(H-E_\alpha)}}{i\omega_m-H+E_\alpha} 
h_i^\dagger 
\right| \gamma' \right\rangle 
\left\langle \gamma' \left|a_j^\dagger  h_j \right| \alpha \right\rangle . 
\label{jj6} 
\end{eqnarray} 
\end{widetext} 

\begin{figure}[tb]
\protect 
\includegraphics[width=7.2cm,clip=]{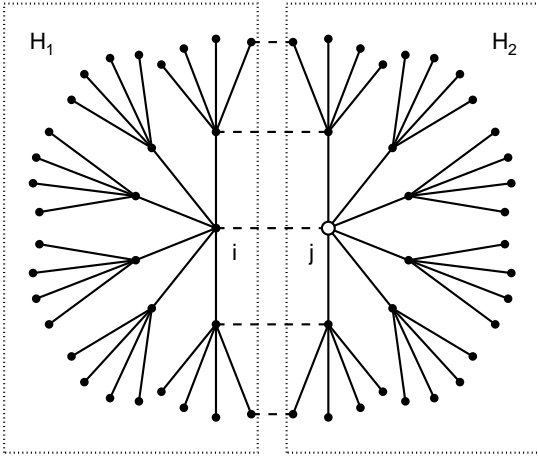} 
\caption{Schematic picture of the hole dynamics 
in the retraceable path approximation for 
$z \rightarrow \infty$. Given a hole at site $j$ [subspace (2)], 
the probability to affect the subspace (1), through the link 
$\langle i, j \rangle$ or other links (dashed connections), 
is $O(1/\sqrt{z})$. 
In the leading order $1/z \rightarrow 0$ the subspace 
$H_1$, $H_2$ are thus independent.} 
\label{f-sketch} 
\end{figure}

Let us now ask ourselves the following question: 
how much does the one-hole state $|\alpha \rangle$ differ from the 
``free-like'' state $|\gamma \rangle$ in the absence of holes? 
It is clear that in the $t=0$ case only the phonon-spin configuration 
on the site $j$ is affected by the presence of the hole, whereas 
all the other sites would be unaffected. 
In the presence of hole dynamics ($t\neq 0$), however, all the 
other sites are in principle affected. 
We remind also that, because of the classical magnetic background, 
the hole dynamics obeys a retraceable path approximation 
just as in a Bethe lattice, as depicted in Fig. \ref{f-sketch}. 
Let us consider now a given specific 
link  of two nearest neighbor sites, $\langle i,j \rangle$. 
In a true Bethe lattice, the whole system can be divided in two subspaces, 
(1) and (2), connected by the hopping term $t_{ij} \propto t/\sqrt{z}$. 
This means that, to leading order, the presence of one hole 
on the site $j$ would not affect the subspace (1). 
Similar considerations hold true for a generic lattice 
under the retraceable path conditions:  additional links 
between the two subspaces (dashed lines in Fig. \ref{f-sketch}), 
will contribute only to $O(1/\sqrt{z})$ 
and they can be neglected in the $z\rightarrow \infty$ limit. 
For practical purposes we can thus split the total 
Hamiltonian as $H=H_1+H_2$, where $H_1$ accounts for the 
phonon-spin degrees of freedom 
of the subspace (1) (not including $j$), while $H_2$ 
contains the phonon-spin degrees of freedom 
of the subspace (2) (not including $i$). 
In a similar way the eigenstates $|\alpha \rangle$, $|\gamma \rangle$ 
can be written (in the leading order of a $1/z$ expansion) as 
$|\alpha \rangle=|\alpha_1 \rangle \otimes |\alpha_2 \rangle$ 
and 
$|\gamma \rangle=|\gamma_1 \rangle \otimes |\gamma_2 \rangle$. 
Employing these results, and noting that 
$\langle \alpha | h_j^\dagger a_j | \gamma \rangle= 
\langle \alpha_1 | h_j^\dagger a_j | \gamma_1 \rangle 
\delta{\alpha_2,\gamma_2}$, 
$\langle \gamma' | a_j^\dagger h_j | \alpha \rangle= 
\langle \gamma'_1 | a_j^\dagger h_j | \alpha_1 \rangle 
\delta{\gamma'_2,\alpha_2}$, 
$\langle \gamma | h_i \ldots h_i^\dagger | \gamma' \rangle= 
\langle \gamma_2 | h_i \ldots h_i^\dagger | \gamma'_2 \rangle 
\delta{\gamma_1,\gamma'_1}$, 
we have thus: 
\begin{widetext} 
\begin{eqnarray} 
\Pi_0(i\omega_m) 
&=& 
\frac{t^2 \mbox{e}^{\beta \mu}}{4z} 
\left\{ 
\sum_{\langle j \rangle_i,\alpha_1,\gamma_1} 
\frac{\mbox{e}^{-\beta E_{\alpha_1}}}{Z^{\rm T}_1} 
\left| 
\left\langle \alpha_1 \left| h_j^\dagger a_j \right| \gamma_1 \right\rangle 
\right|^2 
\right\} 
\nonumber\\
&&\times
\left\{ 
\sum_{i,\alpha_2} 
\frac{\mbox{e}^{-\beta E_{\alpha_2}}}{Z^{\rm T}_2} 
\left\langle \alpha_2 \left| 
h_i 
\frac{1-\mbox{e}^{-\beta(E_{\gamma_1}+H_2-E_{\alpha_1}-E_{\alpha_2})}} 
{i\omega_m-E_{\gamma_1}-H_2+E_{\alpha_1}+E_{\alpha_2}} 
h_i^\dagger 
\right| \alpha_2 \right\rangle 
\right\}, 
\label{jj7} 
\end{eqnarray} 
where $Z^{\rm T}_1$, $Z^{\rm T}_2$ are the partition functions 
of the corresponding subspaces and $\sum_{\langle j \rangle_i}$ denotes 
a sum over the $z$ nearest neighbors of the site $i$. 
Performing the analytical continuation $i\omega_m \rightarrow 
\omega+i\delta$, and 
introducing once more appropriate $\delta$-functions, 
we end up thus with: 
\begin{eqnarray} 
\sigma_0(\omega;n_h)
&=& 
-\frac{\mbox{Im}\Pi_0(\omega+i\delta)}{\omega} 
\nonumber\\
&=& 
\frac{t^2\pi \mbox{e}^{\beta \mu} \left[1-\mbox{e}^{-\beta\omega}\right]} 
{4 z \omega} 
\int d\Omega 
\mbox{e}^{-\beta \Omega} 
\left\{ 
\sum_{\langle j \rangle_i,\alpha_1,\gamma_1} 
\frac{\mbox{e}^{-\beta E_{\gamma_1}}}{Z^{\rm T}_1} 
\left| 
\left\langle \alpha_1 \left| h_j^\dagger a_j \right| \gamma_1 \right\rangle 
\right|^2 
\delta(\Omega-E_{\alpha_1}+E_{\gamma_1}) 
\right\} 
\nonumber\\ 
&&\times 
\left\{ 
\sum_{i,\alpha_2,\lambda_2} 
\frac{\mbox{e}^{-\beta E_{\alpha_2}}}{Z^{\rm T}_1} 
\left|\left\langle \alpha_2 \left| h_i 
\right|\lambda_2\right\rangle\right|^2 
\delta(\omega+\Omega-E_{\lambda_2}+E_{\alpha_2}) 
\right\} 
\nonumber\\ 
&=& 
\frac{t^2\pi N \mbox{e}^{\beta \mu} \left[1-\mbox{e}^{-\beta\omega}\right]} 
{4\omega} 
p_0 p_1\int d\Omega 
\mbox{e}^{-\beta \Omega} 
\bar{\rho}_1(\Omega)\bar{\rho}_0(\omega+\Omega), 
\label{pi0} 
\end{eqnarray} 
\end{widetext} 
where $| \lambda_2 \rangle$ are eigenstates in the $N_h=1$ subspace, 
$p_0=1-x$, $p_1=x$ are the statistical probabilities 
to have no spin defect and one spin defect, respectively, 
and where 
\begin{eqnarray} 
\bar{\rho}_1(\omega)= 
\sum_{\alpha,\gamma} \frac{\mbox{e}^{-\beta E_\gamma}}{Z^{\rm T}} 
\left| \left\langle \alpha  \left| 
h_i^\dagger a_i \right| \gamma \right\rangle \right|^2 
\delta(\omega-E_\alpha+E_\gamma). 
\label{rho1} 
\end{eqnarray} 

The same derivation can be now employed for the case when 
a spin defect is present on the site $i$. 
After few straightforward calculations, we obtain: 
\begin{eqnarray} 
\sigma_1(\omega;n_h) 
&=& 
\frac{t^2\pi N \mbox{e}^{\beta \mu} \left[1-\mbox{e}^{-\beta\omega}\right]} 
{4\omega } 
p_0 p_1 
\nonumber\\ 
&& 
\times 
\int d\Omega 
\mbox{e}^{-\beta \Omega} 
\bar{\rho}_0(\Omega)\bar{\rho}_1(\omega+\Omega). 
\label{pi1} 
\end{eqnarray} 
Summing the two contributions (\ref{pi0}), (\ref{pi1}), 
and reminding $\bar{\rho}_0(\omega)=\bar{\rho}(\omega)$, 
$\bar{\rho}_1(\omega)=\bar{\rho}(\omega+\tilde{J}/2)$, 
after a change of variable we obtain: 
\begin{eqnarray} 
& &\sigma(\omega;n_h) 
= 
\sigma_0(\omega;n_h)+\sigma_1(\omega;n_h) 
\nonumber\\ 
&=& 
\frac{t^2\pi N\mbox{e}^{\beta \mu} \left[1-\mbox{e}^{-\beta\omega}\right]p_0} 
{4\omega } 
\int d\Omega 
\mbox{e}^{-\beta \Omega}\bar{\rho}(\Omega) 
\nonumber\\ 
&& 
\times 
\left[p_1\bar{\rho}(\omega+\Omega+\tilde{J}/2) 
+ p_0 \bar{\rho}(\omega+\Omega-\tilde{J}/2)\right], 
\label{pit} 
\end{eqnarray} 
where we made use also of the relation 
$p_1 \mbox{e}^{\beta \tilde{J}/2}=p_0$. 
Finally, 
dividing Eq. (\ref{pit}) by (\ref{nfin}) we obtain the dimensionless 
quantity Eq. (\ref{sigma}). 
Note that, dividing  Eq. (\ref{pit}) by (\ref{nfin}), 
the common  factors $N \mbox{e}^{\beta \mu}$ 
cancel out, so that the limit $\mu\rightarrow -\infty$ is well defined.


\begin{thebibliography}{99}
\bibitem{notagaux}
We define the amplitude of the Gaussian filter as
three times the variance $\sigma$, so that
$\sigma=\omega_0/5$.
\end{thebibliography}
\end{document}